\documentclass[11pt]{article}
\pdfoutput=1
\usepackage[utf8x]{inputenc}
\usepackage{amssymb,amsmath,mathrsfs,enumerate}
\usepackage{graphicx,rotate,multicol, multirow}
\usepackage{cite}
\usepackage{braket}
\usepackage[normalem]{ulem}
\usepackage{wrapfig}
\usepackage[textfont=it,labelfont=bf]{caption}
\usepackage{bm}
\usepackage[colorlinks=true,
linkcolor=red,
urlcolor=blue,
citecolor=blue]{hyperref}
\usepackage{lineno}
\usepackage{color}
\usepackage{bbold}
\long\def\rpl#1!!#2!!{\textcolor{red}{#1} \textcolor{blue}{#2}}

\makeatother

\newcommand{\bea}{\begin{eqnarray}}
\newcommand{\eea}{\end{eqnarray}}

\newcommand{\be}{\begin{equation}}
\newcommand{\ee}{\end{equation}}
\def\beq{\begin{equation}}
\def\eeq{\end{equation}}
\newcommand{\ba}{\begin{eqnarray}}
\newcommand{\ea}{\end{eqnarray}}
\def\ifmath#1{\relax\ifmmode #1\else $#1$\fi}

\evensidemargin=\oddsidemargin
\def\Eqn#1{Eq.\ (\ref{#1})}
\def\Eqs#1#2{Eqs.\ (\ref{#1}) and (\ref{#2})}
\def\Sect#1{Sec.\,\ref{#1}}
\allowdisplaybreaks

\title{\Large\bf 
	Fingerprinting the Type-Z three Higgs doublet models
}

\author{
	\sf 
	Rafael Boto$^{a,}$\footnote{rafael.boto@tecnico.ulisboa.pt},
	Dipankar Das$^{b,}$\footnote{d.das@iiti.ac.in},
	Luis Lourenco$^{a,}$\footnote{luis.lourenco.1510@tecnico.ulisboa.pt},
	Jorge C. Romão$^{a,}$\footnote{jorge.romao@tecnico.ulisboa.pt},
	Joao P. Silva$^{a,}$\footnote{jpsilva@cftp.ist.utl.pt}
	\\[3mm]
	\small\em
	$^a$ Centro de F\'isica Te\'orica de Part\'iculas-CFTP and Departamento de
	F\'isica,  Instituto Superior T\'ecnico,\\  \small\em
	Universidade de Lisboa, Av
	Rovisco Pais, 1, P-1049-001 Lisboa, Portugal \\ 
	\small\em
	$^b$Indian Institute of Technology (Indore), Khandwa Road, Simrol,
	Indore 453 552, India
}

\date{}

\begin{document}
	
	
\maketitle
\renewcommand*{\thefootnote}{\arabic{footnote}}
\setcounter{footnote}{0}
	
\begin{abstract}
There has been great interest in a model with three Higgs doublets
in which fermions with a particular charge couple to a single and distinct
Higgs field.
We study the phenomenological differences between the two
common incarnations of this so-called Type-Z 3HDM. We point out that the differences
between the two models arise from the scalar potential only. Thus we focus
on observables that involve the scalar self-couplings.
We find it difficult to uncover
features that can uniquely set apart the $Z_3$ variant of the model.
However, by studying the
dependence of the trilinear Higgs couplings on the nonstandard masses, we
have been able to isolate some of the exclusive indicators for the $Z_2\times Z_2$
version of the Type-Z 3HDM.
This highlights the importance of precision measurements of
the trilinear Higgs couplings.
\end{abstract}
	
	\maketitle
	
\section{Introduction}
\label{s:intro}
The Standard Model~(SM) of particle physics has been immensely successful
in describing the electroweak interaction with great precision. However issues
like neutrino mass and dark matter serve as major motivators to look for physics
beyond the SM~(BSM). Very often, such BSM theories extend the minimal scalar
sector of the SM, which consists of only one Higgs-doublet. Therefore, quite
naturally, scalar extensions of the SM are routinely investigated in the
literature. Among these, multi Higgs-doublet models might be the most ubiquitous,
primarily because such extensions preserve the tree-level value of the electroweak
$\rho$-parameter. The simplest extension in this category is the two Higgs-doublet
model which have been studied extensively\cite{Branco:2011iw}. Of late there has been a rise in
interest in the study of three Higgs-doublet models~(3HDMs)\cite{Keus:2013hya,Ivanov:2012fp} where, as the
name suggests, the scalar sector contains three Higgs-doublets.

In the studies of multi Higgs-doublet models it is very often assumed that fermions of a
particular charge couple to a single scalar doublet. This will make fermion mass matrices
proportional to the corresponding Yukawa matrices and diagonalization of the mass
matrices will automatically ensure the simultaneous diagonalization of the Yukawa matrices
as well. As a result the model will be free from scalar mediated flavor changing neutral
couplings~(FCNCs) at the tree-level. In Ref.~\cite{Ferreira:2010xe} it was explicitly
demonstrated that tree-level FCNCs are absent if and only if there is a basis for the Higgs doublets in which
all the fermions of a given electric charge couple to only one Higgs doublet.
Such an aspect of the model is quite desirable in
view of the flavor data\cite{ParticleDataGroup:2022pth}. These types of constructions are usually referred to
as models with natural flavor conservation~(NFC)~\cite{Glashow:1976nt} in the literature,
of which there are five independent possibilities.

Following the terminologies of Ref.~\cite{Yagyu:2016whx},
 one can entertain four types of flavor universal NFC
models, namely Type-I, Type-II, Type-X, and Type-Y, within the 2HDM framework.
All these Yukawa structures have been concisely summarized
in Table~\ref{t:NFC}. 
Beyond these four options, there is one more interesting possibility where a particular scalar doublet is reserved exclusively for each type of massive fermion. This implies that the up-type quarks, the down-type quarks, and the charged-leptons couple to separate scalar doublets.  
Evidently, such an arrangement of Yukawa couplings is impossible within a 2HDM framework and one needs at least three scalar doublets to accommodate it.  
In this paper, we will refer to this possibility as the `Type-Z Yukawa' and subsequently, the 3HDMs that feature a Type-Z Yukawa structure will be collectively called `Type-Z 3HDMs'. 
These Type-Z 3HDMs have gained a lot of attention in the recent past.  
Theoretical constraints from unitarity and boundedness from below (BFB) have been studied in Refs.~\cite{Bento:2022vsb, Boto:2022uwv, Moretti:2015cwa}, the alignment limit is analyzed in Refs.~\cite{Das:2019yad, Pilaftsis:2016erj}, the custodial limit has been studied
in Refs.~\cite{Das:2022gbm},
and quite recently, the phenomenological analysis involving the
flavor and Higgs data have been performed in Refs.~\cite{Chakraborti:2021bpy, Boto:2021qgu}.
Other related studies appear in
\cite{Cree:2011uy, Akeroyd:2016ssd, Alves:2020brq, Logan:2020mdz}.

There are usually two different ways in which a Type-Z Yukawa structure is realized.
The first method employs a $Z_3$ symmetry\cite{Das:2019yad} whereas the second option uses
a $Z_2\times Z_2$ symmetry\cite{Akeroyd:2016ssd}. Our objective in this paper will be to point out observable
features which can distinguish between the two avatars of Type-Z 3HDMs. Since the Yukawa sector
in both versions of Type-Z 3HDM is identical, we will turn our attention to the scalar potential
with the hope that some distinguishing aspects can be uncovered. As we will see,
only some of the quartic terms in the scalar potential mark the difference between
the two variants of Type-Z 3HDM. We will therefore focus on the theoretical constraints
from unitarity and BFB which concern the quartic parameters of the scalar potential.
We hope that these constraints, in particular, will impact the parameter space
in the scalar sector differently for the two Type-Z models. As a result, we
expect to encounter some practical distinguishing features of these two models.

Our article will be organized as follows. In Sec.~\ref{s:model} we will outline the two
different options for obtaining a Type-Z Yukawa structure along with the corresponding
implications for the scalar potential. In Sec.~\ref{s:constraints} we list the different
constraints (both theoretical and phenomenological) faced by the scalar sectors of the 3HDMs
under consideration. In Sec.~\ref{s:results} we spell out the details of our numerical analysis
and highlight the important outcomes. We summarize our findings and draw our conclusions
in Sec.~\ref{s:conclusions}.

\section{The model}
\label{s:model}
We have already presented the notion of NFC in the introduction. There are a few different ways of
obtaining NFC in a 3HDM framework, which have been listed in a concise manner in Table~\ref{t:NFC} where $\phi_1$, $\phi_2$ and $\phi_3$ represent the three
Higgs-doublets that constitute the scalar sector of our model.
Among these, we are particularly interested in the possibility of Type-Z Yukawa structure which requires
a 3HDM scalar sector at the very least. There are two different ways to ensure a Type-Z Yukawa structure.
The first option is to employ a $Z_3$ symmetry as follows:
\begin{subequations}
\label{e:sym}
\begin{eqnarray}
\label{e:Z3sym}
\phi_1 \to e^{2\pi i/3} \, \phi_1 \,, \qquad  \phi_2 \to e^{4\pi i/3} \phi_2 \,,  \qquad	\ell_R \to e^{4\pi i/3} \ell_R \,,  \qquad  d_R \to e^{2\pi i/3} \,d_R \,,
\end{eqnarray}
and the second option will be to use a $Z_2\times Z_2^\prime$ symmetry in the following manner:
\begin{eqnarray}
\label{e:Z2sym}
Z_2:&&\phi_1 \to -\phi_1 \, , \quad  \ell_R \to -\ell_R \,, \\
Z'_2:&&\phi_2 \to -\phi_2 \, , \quad  d_R \to -d_R \,.
\end{eqnarray}
\end{subequations}
In the equations above the down-type quark and charged-lepton right-handed fields are denoted as $d_R$ and $\ell_R$, respectively. Since both the symmetries in \Eqn{e:sym} entail the same
Type-Z Yukawa couplings, we must turn our attention to the scalar sector phenomenologies for possible
distinguishable features. The symmetries in \Eqn{e:sym} would obviously have their repercussions
on the 3HDM scalar potential. To this end we note that the scalar potentials in both these cases consist
of a common part as follows:
\begin{subequations}
\label{e:common}
\begin{eqnarray}
	V_C &=& V_2 + V_{4C} \,, \qquad {\rm where,} \\
	V_2 &=& m_{11}^2\phi_1^\dagger\phi_1 + m_{22}^2\phi_2^\dagger\phi_2 + m_{33}^2\phi_3^\dagger\phi_3 -
	\left[m_{12}^2(\phi_1^\dagger\phi_2) +m_{13}^2(\phi_1^\dagger\phi_3)
	+m_{23}^2(\phi_2^\dagger\phi_3)+\text{h.c.}\right] \,, \label{e:V2} \\
	V_{4C} &=& \lambda_1(\phi_1^\dagger\phi_1)^2 +\lambda_2(\phi_2^\dagger\phi_2)^2
	+\lambda_3(\phi_3^\dagger\phi_3)^2+ \lambda_4(\phi_1^\dagger\phi_1)(\phi_2^\dagger\phi_2)
	+\lambda_5(\phi_1^\dagger\phi_1)(\phi_3^\dagger\phi_3)\nonumber \\ 
	&& +\lambda_6(\phi_2^\dagger\phi_2)(\phi_3^\dagger\phi_3)	+\lambda_7(\phi_1^\dagger\phi_2)(\phi_2^\dagger\phi_1)
	+\lambda_8(\phi_1^\dagger\phi_3)(\phi_3^\dagger\phi_1)	+\lambda_9(\phi_2^\dagger\phi_3)(\phi_3^\dagger\phi_2) \,.
	\label{e:V4} 
\end{eqnarray}
\end{subequations}
\begin{table}
	\centering
	\begin{tabular}{ |c|c ccc|c| } 
		\hline
		fermion type & Type-I & Type-II & Type-X & Type-Y & Type-Z \\ 
		\hline
		\hline
	up quarks ($u$) & $\phi_3$ & $\phi_3$ & $\phi_3$ & $\phi_3$ & $\phi_3$  \\ 
	down quarks	($d$) & $\phi_3$ & $\phi_2$ & $\phi_3$ & $\phi_2$ & $\phi_2$  \\ 
	charged leptons	($\ell$) & $\phi_3$ & $\phi_2$ & $\phi_2$ & $\phi_3$ & $\phi_1$ \\
		\hline
	\end{tabular}
	\caption{\small Distinct possibilities for NFC in a 3HDM framework. The first four types can also be obtained within 2HDMs but the Type-Z requires at least a 3HDM. In our convention, the scalar doublet coupling to the up-type quarks is always labeled as $\phi_3$.
	}
	\label{t:NFC}
\end{table}
Note that in the expression for $V_2$, we have allowed terms that softly break the symmetries defined in
\Eqn{e:sym}. These will be important if we wish to access arbitrarily heavy nonstandard scalars
(decoupled from physics at the electroweak scale) without spoiling perturbative unitarity\cite{Bhattacharyya:2014oka,Carrolo:2021euy,Faro:2020qyp}.
The differences between the symmetries in \Eqs{e:Z3sym}{e:Z2sym} are captured by the following
quartic terms in the scalar potential:
\begin{subequations}
\label{e:VZ2Z3}
\begin{eqnarray}
	V_{Z3} &=& V_C + \left[\lambda_{10}(\phi_1^\dagger\phi_2)(\phi_1^\dagger\phi_3)
	+\lambda_{11}(\phi_1^\dagger\phi_2)(\phi_3^\dagger\phi_2) +\lambda_{12}(\phi_1^\dagger\phi_3)(\phi_2^\dagger\phi_3)+
	\text{h.c.}\right] \,,
	\label{e:VZ3} \\
	V_{Z2} &=& V_C  +\left[\lambda'_{10}(\phi_1^\dagger\phi_2)^2 +
	\lambda'_{11}(\phi_1^\dagger\phi_3)^2 +	\lambda'_{12}(\phi_2^\dagger\phi_3)^2 +	\text{h.c.}\right]\,.
	\label{e:VZ2}
\end{eqnarray}
\end{subequations}
We, therefore, hope to find distinguishing aspects of these models by tracking the effects of these
additional terms.

In order to do this, it is important to conveniently parametrize our models in terms
of the physical masses and mixings. We will closely follow the notations and conventions
of some earlier works\cite{Das:2022gbm,Boto:2021qgu}. However, for the sake of completeness, we will give
a brief summary of the important expressions which will be crucial for our numerical
analysis later. To begin with, let us write the $k$-th scalar doublet, after spontaneous
symmetry breaking, as follows:
\begin{eqnarray}
\label{e:phik}
\phi_k=\frac{1}{\sqrt{2}}\begin{pmatrix}
\sqrt{2} w_k^+ \\
v_k+h_k+i z_k
\end{pmatrix} \, ,
\end{eqnarray}
where $v_k$ is the vacuum expectation value~(VEV) of $\phi_k$, assumed to be real. The three VEVs, $v_1$, $v_2$
and $v_3$ are conveniently parametrized as
\begin{eqnarray}
\label{e:vevs}
v_1 = v \cos\beta_1 \cos\beta_2 \,, \quad v_2 = v \sin\beta_1 \cos\beta_2 \,, \quad
v_3 = v \sin\beta_2\,,
\end{eqnarray}
where $v = \sqrt{v_1^2+v_2^2+v_3^2}$ is the total electroweak (EW) VEV.
The component fields in \Eqn{e:phik} will mix together and will give rise to
two pairs of charged scalars ($H_{1,2}^\pm$), two physical pseudoscalars ($A_{1,2}$)
 and three CP-even neutral scalars ($h, H_{1,2}$).\footnote{ We are implicitly assuming 
 CP conservation in the scalar sector so that such a classification of the physical
scalar spectrum is possible.}
For the charged and pseudoscalar sectors, the physical scalars can be obtained via the following $3\times3$ rotations,
\begin{equation}
\label{e:charged}
\begin{pmatrix} \omega^\pm\\ H_1^\pm \\ H_2^\pm\end{pmatrix} =
\mathcal{O}_{\gamma_2}\mathcal{O}_\beta
\begin{pmatrix} w_1^\pm\\ w_2^\pm \\ w_3^\pm\end{pmatrix}\; , \quad
\quad
\begin{pmatrix} \zeta\\ A_1 \\ A_2\end{pmatrix} =
\mathcal{O}_{\gamma_1} \mathcal{O}_\beta 
\begin{pmatrix} z_1\\ z_2 \\ z_3\end{pmatrix} ,
\end{equation}
where, the rotation matrices are given by
\begin{eqnarray}\label{eq:gammaRots}
{\cal O}_{\gamma_1} =
\begin{pmatrix}
1 & 0 & 0 \\
0 & \cos\gamma_1 & -\sin\gamma_1 \\
0 & \sin\gamma_1 & \cos\gamma_1 \end{pmatrix}  \label{e:Ogamma1} \,,
\quad{\cal O}_{\gamma_2} =
\begin{pmatrix}
1 & 0 & 0 \\
0 & \cos\gamma_2 & -\sin\gamma_2 \\
0 & \sin\gamma_2 & \cos\gamma_2 \end{pmatrix}  \label{e:Ogamma2} \, ,
\end{eqnarray}
and
\begin{eqnarray}\label{eq:betaRot}
{\cal O}_{\beta} =
\begin{pmatrix} \cos\beta_2 \cos\beta_1 & \cos\beta_2 \sin\beta_1
&  \sin\beta_2 \\
-\sin\beta_1 & \cos\beta_1  &  0  \\
-\cos\beta_1 \sin\beta_2 & -\sin\beta_1\sin\beta_2 & \cos\beta_2 
\end{pmatrix}.
\end{eqnarray}
In \Eqn{e:charged}, $\omega^\pm$ and $\zeta$ stand for the charged and the neutral
Goldstone fields respectively.
For the CP-even sector, we can obtain the physical scalars as follows:
\begin{eqnarray}
\label{e:ab3hdm}
\begin{pmatrix}
h \\
H_1 \\
H_2 
\end{pmatrix}
&=&{\cal O}_ \alpha 
\begin{pmatrix}
h_1 \\
h_2 \\
h_3 
\end{pmatrix}
\label{CP-even}
\end{eqnarray}
where
\begin{subequations}
	\label{e:Oa}
	\begin{eqnarray}
	{\cal O}_\alpha &=& {\cal R}_3 \cdot  {\cal R}_2\cdot {\cal R}_1 \,,
	\end{eqnarray}
	with
	\begin{equation}
	\label{e:R}
	{\cal R}_1 = \begin{pmatrix}
	\cos \alpha_1 & \sin \alpha_1 & 0 \\
	-\sin \alpha_1 & \cos \alpha_1 & 0 \\
	0 & 0 & 1 \end{pmatrix}
	\,, \quad {\cal R}_2 = \begin{pmatrix}
	\cos \alpha_2 & 0 & \sin \alpha_2  \\
	0 & 1 & 0 \\
	-\sin \alpha_2 & 0 & \cos \alpha_2 
	\end{pmatrix}\,,  \quad
	{\cal R}_3 = \begin{pmatrix}
	1 & 0 & 0 \\
	0 & \cos \alpha_3 &  \sin \alpha_3  \\
	0 & -\sin \alpha_3 & \cos \alpha_3 
	\end{pmatrix}.
	\end{equation}
\end{subequations}
Of course, these physical masses and mixings cannot be completely arbitrary as
they will have to negotiate a combination of theoretical and phenomenological
constraints which will be described in the next section.


\section{Constraints}
\label{s:constraints}
In this section we study the constraints that must be applied to the
model parameters in order to ensure theoretical and phenomenological consistency.
On the phenomenological side, we first need to guarantee the presence of
a SM-like Higgs which will be identified with the scalar boson discovered
at the LHC. This can be easily accommodated by staying close to the
`alignment limit'\cite{Das:2019yad} in 3HDM, defined by the condition
\begin{eqnarray}
	\alpha_1 = \beta_1 \,, \qquad
	\alpha_2 = \beta_2 \,.
\end{eqnarray}
In this limit, the lightest CP-even scalar, $h$, will possess the exact SM-like
couplings at the tree-level and constraints from the Higgs signal strengths will
be trivially satisfied. However, we will more interested in the extent of deviation
from the exact alignment limit allowed from the current measurements of the Higgs
signal strengths\cite{ATLAS:2022vkf}.
We then define the Higgs signal strength as follows:
\begin{eqnarray}
	\mu_i^f =\left(\frac{\sigma_i^{\text{3HDM}}(pp\to h) }{\sigma_i^{\text{SM}}(pp\to h)}\right)\left(\frac{\text{BR}^{\text{3HDM}}(h\to f)}{\text{BR}^{\text{SM}}(h\to f)}\right) \,,
	\label{e:ss}
\end{eqnarray}
where the subscript `$i$' denotes the production mode and the superscript `$f$'
denotes the decay channel of the SM-like Higgs scalar. Starting from the collision of two protons, the relevant production
mechanisms include gluon fusion~($ggF$), vector boson fusion~($VBF$),
associated production with a vector boson ($VH$, $V = W$ or $Z$), and
associated production with a pair of top quarks ($ttH$). The SM cross
section for the gluon fusion process is calculated using HIGLU
\cite{Spira:1995mt}, and for the other production mechanisms we use
the prescription of Ref.~\cite{deFlorian:2016spz}.

Next we need to satisfy the constraints arising from the electroweak $S$, $T$ and
$U$ parameters. We will use the analytic expressions derived in Ref.~\cite{Grimus:2007if}
and compare them with the corresponding fit values given in Ref.~\cite{Baak:2014ora}.
It might be worth pointing out that, similar to the 2HDM case, one can easily leap
over the $T$-parameter constraints by requiring\cite{Das:2022gbm}
\begin{eqnarray}
m_{C1} = m_{A1} \,, \qquad m_{C2} = m_{A2} \,, \qquad \gamma_1 =\gamma_2 \,.
\end{eqnarray}

We also take into consideration the bounds coming from flavor data. In the Type-Z
3HDM there are no FCNCs at the tree-level. Therefore the only NP contribution at
one-loop order to observables such as $b\to s\gamma$ and the neutral meson mass
differences will come from the charged scalar Yukawa couplings. It was found in
Ref.~\cite{Chakraborti:2021bpy} that the constraints coming from the meson mass
differences tend to exclude very low values of $\tan\beta_{1,2}$. Therefore, we
only consider
\begin{eqnarray}
	\tan\beta_{1,2} > 0.3 \,,
\end{eqnarray}
to safeguard ourselves from the constraints coming from the neutral meson mass
differences. To deal with the constraints stemming from  $b\to s\gamma$, we
follow the procedure described in Refs.~\cite{Florentino:2021ybj,Boto:2021qgu, Akeroyd:2020nfj}
and impose the following restriction
\begin{equation}
\label{e:b2sg}
2.87 \times 10^{-4} < \text{BR}(B\to X_s \gamma) < 3.77 \times 10^{-4}\,,
\end{equation}
which represents the $3\sigma$ experimental limit.

Additionally, we also take into account the bounds from the direct searches
for the heavy nonstandard scalars. For this purpose, we use \texttt{HiggsBounds-5.9.1} following Ref.~\cite{Bechtle:2020pkv}
where a list of all the relevant experimental searches
can be found. It should be noted that
we have allowed for decays with off-shell scalar bosons,
using the method explained in Ref.~\cite{Romao:1998sr}.

For the theoretical constraints, we first ensure the perturbativity of the
Yukawa couplings. For the Type-Z Yukawa structure, the top, bottom, and tau Yukawa couplings are given by 
\begin{eqnarray}\label{eq:yukawa}
y_t = \frac{\sqrt{2}\, m_t }{v \sin\beta_2}\;,
\quad y_b = \frac{\sqrt{2}\, m_b }{v \sin\beta_1 \cos\beta_2} \;,
\quad y_\tau = \frac{\sqrt{2}\, m_\tau }{v \cos\beta_1 \cos\beta_2} \;,
\end{eqnarray}
which follow from our convention that $\phi_3$, $\phi_2$, and $\phi_1$ couple to up-type quarks, down-type quarks, and charged leptons respectively. To maintain the perturbativity of Yukawa couplings, we impose $\lvert y_t \rvert,\lvert y_b\rvert,\lvert y_\tau\rvert < \sqrt{4\pi}$. 
Throughout our paper, we have used values of $\tan\beta_{1,2}$ which are consistent with this perturbative region.

However, we are mainly interested in the effects of the theoretical constraints from
perturbative unitarity and BFB conditions. These constraints directly affect the scalar
potential and therefore can potentially have different implications for the $Z_3$ and
$Z_2\times Z_2$ incarnations of the Type-Z 3HDM. For the  unitarity constraints, we use
the algorithm presented in Refs.~\cite{Bento:2017eti,Bento:2022vsb}.
For the BFB constraints we use only
the sufficient conditions of Ref.~\cite{Boto:2021qgu} for the $ Z_3$ model and the sufficient conditions of Ref.~\cite{Boto:2022uwv} for the $ Z_2\times Z_2$
model.

\section{Analysis and results}
\label{s:results}
In both versions of the Type-Z 3HDM the scalar potential of \Eqn{e:common}
contains a total of 18 parameters\footnote{We are assuming all the parameters to
be real.} (6 bilinear parameters and 12 quartic parameters). For our numerical
analysis, we will trade these 18 parameters in favor of an equivalent but more
convenient set of parameters which have a more direct connection to the physical
reality. As a first step, we use the minimization conditions to replace three
quadratic parameters, $m_{11}^2$, $m_{22}^2$, and $m_{33}^2$, by the three
VEVs, $v_1$, $v_2$, and $v_3$ which, in turn, are further exchanged with $v$,
$\tan\beta_1$, and $\tan\beta_2$. The 12 quartic parameters are purposefully
interchanged with the 7 physical masses (two charged scalar masses labeled 
as $m_{C1}$ and $m_{C2}$, two pseudoscalar masses labeled as
$m_{A1}$ and $m_{A2}$, and three CP-even scalar masses labeled as
$m_h$, $m_{H1}$ and $m_{H2}$) and 5 mixing angles appearing in
\Eqs{eq:gammaRots}{e:Oa}.

	For each of the symmetry constrained 3HDM, we built a dedicated
	code, which is an extension of our previous
	codes~\cite{Fontes:2014xva,Florentino:2021ybj,Boto:2021qgu}.
	We take $v = 246\,\text{GeV}$ and $m_{h} = 125\,\text{GeV}$ as experimental
	inputs.
	The remaining parameters will be randomly scanned within the following
	ranges:~\footnote{More details about \eqref{eq:scanparameters} are given
after \Eqn{e:Al-10} below.}
	\begin{subequations}
	\label{e:scan}
		\begin{eqnarray}
		&& \alpha_1,\, \alpha_2,\, \alpha_3,\, \gamma_1,\, \gamma_2\, \in \left[-\frac{\pi}{2},\frac{\pi}{2}\right];\qquad \tan{\beta_1},\,\tan{\beta_2}\,\in \left[0.3,10\right]; \\
		&& m_{H1}\,,\, m_{H2}\,	\in \left[125,1000\right]\,\text{GeV} \,;\\
		&& 	m_{A1}\,,\,m_{A2}\,,\, m_{C1}\,,\,m_{C2}\,
		\in \left[100,1000\right]\,\text{GeV}\,; \\
		&&	m^2_{12}\,, m^2_{13}\,, m^2_{23} \, \in  \left[- 10^{7}, 10^{7}\right]\,
		\text{GeV}^2\, . \qquad 
		\label{eq:scanparameters}
		\end{eqnarray}
	\end{subequations}
	The lower limits chosen for the nonstandard masses satisfy the
	constraints listed in Ref.~\cite{Aranda:2019vda}
	and the lower limit on $\tan\beta_{1,2}$ enables us to easily evade
	the constraints from the meson mass differences.

	When studying 3HDM, it was
	noted~\cite{Das:2019yad,Boto:2021qgu,Chakraborti:2021bpy} that in
	order to be able to generate good points in an easy way one should not be far away
	from alignment, defined as the situation where the lightest Higgs
	scalar has the SM couplings. It was shown in Ref.~\cite{Das:2019yad}
	that this corresponds to the case when
	\begin{equation}
	\label{e:alignment}
	\alpha_1=\beta_1 \,,\quad \alpha_2=\beta_2 \,, 
	\end{equation}
	with the remaining parameters allowed to be free, although subject to
	the constraints below. It turns out that for $Z_3$ 3HDM~\cite{Boto:2021qgu}, this
	constraint alone is not enough to generate a sufficiently large set of
	good points starting from a completely unconstrained scan as in
	Eq.~(\ref{eq:scanparameters}).
	In  Ref.~\cite{Chakraborti:2021bpy} it was noted that all the theoretical
	and experimental constraints on the scalar sector can be easily negotiated in the
`maximally symmetric limit' of 3HDM\cite{Darvishi:2021txa}. As pointed out in Ref.~\cite{Chakraborti:2021bpy} one can easily migrate to the maximally symmetric limit
by imposing the following relations among the physical parameters:
	%
	\begin{equation}
	\label{eq:max-sym}
	\gamma_1=\gamma_2=-\alpha_3 \,,\quad m_{H1}=m_{A1}=m_{C1}\,, \quad
	m_{H2}=m_{A2}=m_{C2} \,.
	\end{equation}
Additionally, the maximally symmetric limit also requires
 the soft breaking parameters to be related as follows:
	\begin{subequations}
	\label{eq:soft-sym}
	\begin{eqnarray}
		m^2_{12} &=& c_{\beta_{1}}^2 c_{\gamma_{2}} s_{\beta_{2}} s_{\gamma_{2}}
		\left(m_{C1}^2-m_{C2}^2\right)+c_{\beta_{1}} s_{\beta_{1}}
		\left[s_{\beta_{2}}^2 \left(c_{\gamma_{2}}^2 m_{C2}^2+m_{C1}^2
		s_{\gamma_{2}}^2\right)-c_{\gamma_{2}}^2 m_{C1}^2-m_{C2}^2
		s_{\gamma_{2}}^2\right]\nonumber \\
		&& +c_{\gamma_{2}} s_{\beta_{1}}^2 s_{\beta_{2}} s_{\gamma_{2}}
		\left(m_{C2}^2-m_{C1}^2\right) \,, \\[+3mm]
		m^2_{13} &=&-c_{\beta_{2}} \left[c_{\beta_{1}} s_{\beta_{2}} \left(c_{\gamma_{2}}^2
		m_{C2}^2+m_{C1}^2 s_{\gamma_{2}}^2\right)-c_{\gamma_{2}} s_{\beta_{1}}
		s_{\gamma_{2}} \left(m_{C1}^2-m_{C2}^2\right)\right]\,, \\[+3mm]
		m^2_{23} &=& -c_{\beta_{2}} \left[c_{\beta_{1}} c_{\gamma_{2}} s_{\gamma_{2}}
		\left(m_{C1}^2-m_{C2}^2\right)+s_{\beta_{1}} s_{\beta_{2}}
		\left(c_{\gamma_{2}}^2 m_{C2}^2+m_{C1}^2 s_{\gamma_{2}}^2\right)\right]
		\,,
	\end{eqnarray}
	\end{subequations}
where $s_x$ and $c_x$ are shorthands for $\sin x$ and $\cos x$ respectively.
Therefore, we can make our numerical study very efficient by strategically scanning
in the `neighborhood' of \Eqs{eq:max-sym}{eq:soft-sym}.

In a previous phenomenological study of the $Z_3$ version of the
Type-Z 3HDM~\cite{Boto:2021qgu} we found that one can
	deviate from the exact relations of Eqs.~(\ref{e:alignment}),
	(\ref{eq:max-sym}) and (\ref{eq:soft-sym}) 
	by a given percentage (10\%, 20\%, 50\%) thereby enhancing the possibility
	of new BSM signals, while at the same time being able to generate
	adequate number of data points. 
	To exemplify, we can ensure to be within $x\%$ of the alignment
	condition of \Eqn{e:alignment} by choosing to scan within the following
	range:
	\begin{equation}
	\label{Al-1}
	\frac{\alpha_1}{\beta_1} \,,\ 
	\frac{\alpha_2}{\beta_2}\, \in\, [1-x\%\,,1+x\%]\, .
	\end{equation}
Extending this prescription we can simultaneously incorporate \Eqs{e:alignment}{eq:max-sym}
by scanning within the following range:\footnote{The alignment limit \textbf{(Al-x\%)} was
referred to in \cite{Boto:2021qgu} as \textbf{(Al-2-x\%)}.}
	\begin{equation}
	\label{e:Al-20}
	\frac{\alpha_1}{\beta_1},\ 
	\frac{\alpha_2}{\beta_2},\ 
	\frac{\gamma_2}{\gamma_1},\ 
	\frac{-\alpha_3}{\gamma_1},\ 
	\frac{m_{A1}}{m_{H1}},\ 
	\frac{m_{C1}}{m_{H1}},\ 
	\frac{m_{A2}}{m_{H2}},\ 
	\frac{m_{C2}}{m_{H2}}\, \in\, [0.8,1.2]\, .
	\ \ \ \textbf{(Al-20\%)}
	\end{equation}
	The set of points obtained after scanning over the above range
	will be labeled as `Al-20\%' 
	in the subsequent text and plots.
	In a similar manner we generate another data set labeled as 
	`Al-10\%' 
        which are relatively closer to the conditions of
	\Eqs{e:alignment}{eq:max-sym} by scanning over the following range:
	\begin{equation}
	\label{e:Al-10}
	\frac{\alpha_1}{\beta_1},\ 
	\frac{\alpha_2}{\beta_2},\ 
	\frac{\gamma_2}{\gamma_1},\ 
	\frac{-\alpha_3}{\gamma_1},\ 
	\frac{m_{A1}}{m_{H1}},\ 
	\frac{m_{C1}}{m_{H1}},\ 
	\frac{m_{A2}}{m_{H2}},\ 
	\frac{m_{C2}}{m_{H2}}\, \in\, [0.9,1.1]\, .
	\ \ \ \textbf{(Al-10\%)}
	\end{equation}
	In this context it should be noted that the soft-breaking parameters,
	whenever they are free, are scanned in a very similar manner in
	the vicinity of \Eqn{eq:soft-sym}.


\begin{table}[ htbp!]
  \centering
  \makebox[0pt][c]{\parbox{0.96\textwidth}{%
      \hskip -2mm
      \begin{minipage}[b]{0.49\textwidth}\centering
        \begin{tabular}{|c|c|c|c|c|}\hline
          \multicolumn{5}{|c|}{$ Z_2 \times  Z_2$\  (AL-10\%) }\\\hline\hline
          Check   &    N   &  Y     & 100*p   & 100*$\delta_p$\\\hline\hline
          STU  &   500000  &   407162  &   81.432  &    0.172\\*[1mm]
          BFB  &  5000000 &    380066  &    7.601  &    0.013\\*[1mm]
          Unitarity  &   500000  &    26386  &    5.277 &     0.033\\*[1mm]
          $b \rightarrow s \gamma$  &    50000 &     22198  &   44.396 &     0.358\\*[1mm]
           $\mu$'s  &   50000 &  4168     &  8.336  &  0.282   \\\hline
        \end{tabular}
      \end{minipage}
      \hfill
      \begin{minipage}[b]{0.49\textwidth}\centering
        \begin{tabular}{|c|c|c|c|c|}\hline
          \multicolumn{5}{|c|}{$ Z_3$\  (AL-10\%) }\\\hline\hline
          Check    &   N  &  Y   & 100*p   & 100*$\delta_p$\\\hline\hline
          STU  &   500000  &   407176  &   81.435  &    0.172\\*[1mm]
          BFB &   5000000  &    42703  &    0.854  &    0.004\\*[1mm]
          Unitarity  &   500000  &    18424  &    3.685  &    0.027\\*[1mm]
          $b \rightarrow s \gamma$  &    50000  &    21810  &    43.62  &    0.354\\*[1mm]
          $\mu$'s &     50000  &   4141  & 8.282  & 0.271    \\\hline
        \end{tabular}
      \end{minipage}
    }}
\caption{Impact of individual constraints for the two Type-Z models
  while the scanning is done following \Eqn{e:Al-10}.}
  \label{tab:k10}
\end{table}

To explicitly demonstrate the efficiency of our scanning method, we
display in Table~\ref{tab:k10} how restrictive the
individual constraints of \Sect{s:constraints} can be. In these tables,
$N$ represents the number of initial input points and $Y$ stands for
the number of output points that can successfully pass through a given
constraint labeled appropriately.
Thus $p = Y/N$ gives an estimate for the
probability of successfully negotiating a particular constraint. The
quantity $\delta_p$ represents the typical uncertainty associated with
the estimate of $p$ and is calculated using the formula for the
propagation of errors.
From Table \ref{tab:k10} it should be evident that the
BFB constraints have a very low acceptance ratio for the input
points.
We should point out that
our choice of scanning around \Eqs{e:alignment}{eq:max-sym} does
definitely increase the number of output points that pass
through all the constraints.
This is detailed in the Appendix,
where we show equivalent numbers for a run generating points within
50\% of the alignment limit.

Now that our scan strategy has been laid-out clearly, we can proceed
to describe the results from our numerical studies. We will do this in
two stages. At first we will demonstrate the results from the general
scans and point out features that may distinguish between the two
variants of Type-Z 3HDMs.  In the second part we will presume that
some nonstandard scalars have been discovered and therefore we will
work with some illustrative benchmark points in the hope of making the
distinction between the two models more pronounced.

\begin{figure}[htbp!]
  \centering
\includegraphics[width=0.47\textwidth]{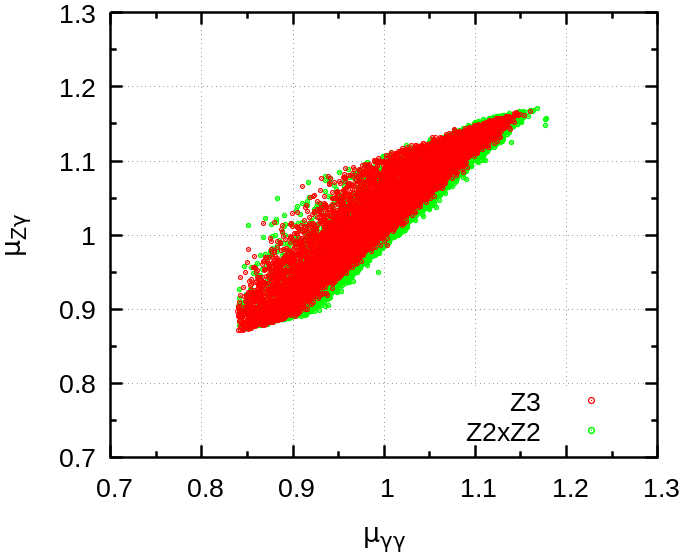}
\caption{Output points that pass all the constraints are plotted in
  $\mu_{\gamma\gamma}$ vs $\mu_{Z\gamma}$ 
  plane for the gluon fusion production channel. The scanning is
  done assuming the Al-10\% condition of \Eqn{e:Al-10}.
  The red and the green points correspond to the
  $Z_3$ and $ Z_2 \times Z_2$ models respectively.}
\label{fig:mugagak10}
\end{figure}
We have to always keep in mind that the difference between the two versions of
Type-Z 3HDM is marked by the scalar potential. Therefore, we focus on the
measurements that involve the scalar self-couplings. Quite naturally, our first
choice will be to study $\mu_{\gamma\gamma}$ and $\mu_{Z\gamma}$ (Higgs signal
strengths in the two photon and $Z$-photon channels respectively) which pick up
extra contributions from charged scalar loops that depend on couplings of the
form $hH_i^+ H_i^-$ ($i=1,2$). However, as we have displayed
in Fig.~\ref{fig:mugagak10}, the points that pass through all the constraints
span very similar regions in the $\mu_{\gamma\gamma}$ vs $\mu_{Z\gamma}$
plane for both versions of Type-Z 3HDM. Thus no significant distinction between
the two models can be made from $\mu_{\gamma\gamma}$ and $\mu_{Z\gamma}$.


\begin{figure}[htbp!]
	\centering
	\includegraphics[width=0.31\textwidth]{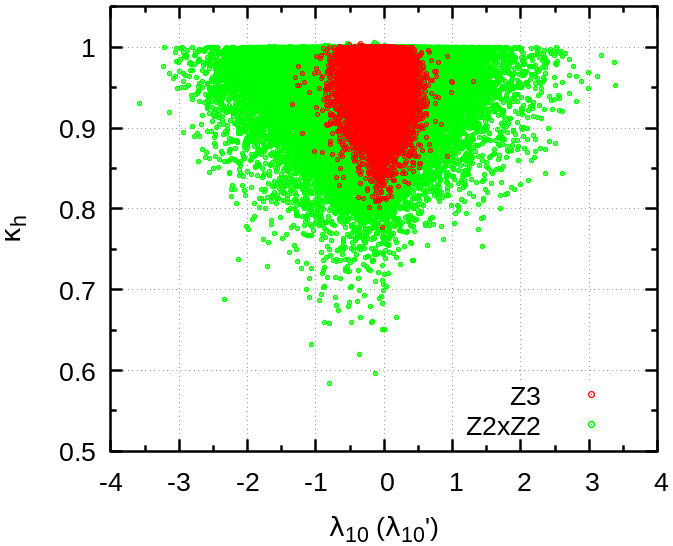}
	\includegraphics[width=0.31\textwidth]{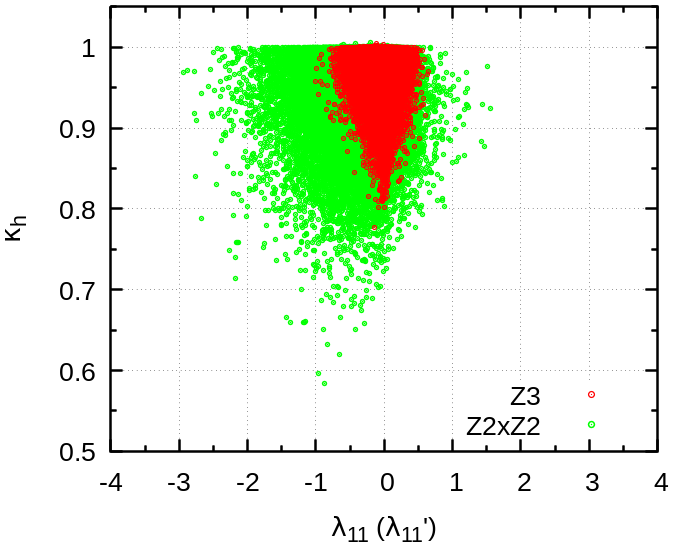}
	\includegraphics[width=0.31\textwidth]{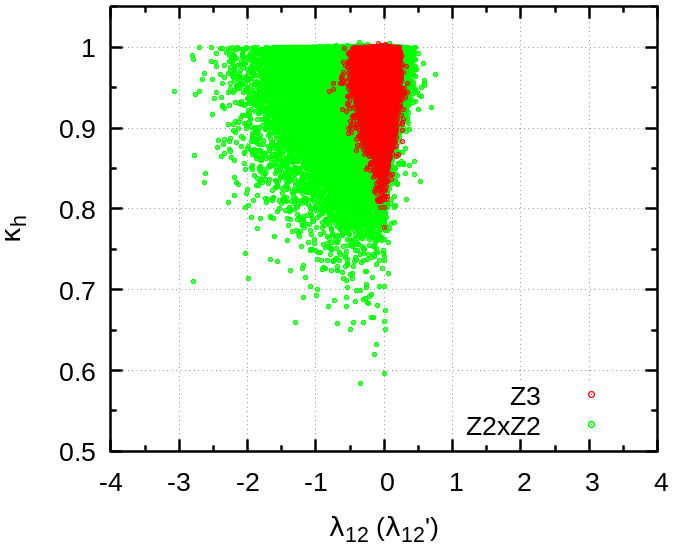}
	\caption{Points that pass through all the constraints are plotted in the
	$\kappa_h$ vs $\lambda_{10}^{(\prime)}, \lambda_{11}^{(\prime)}, \lambda_{12}^{(\prime)}$ plane. The scanning is
	done assuming the Al-10\%  condition of \Eqn{e:Al-10}.
	The red and the green points correspond to the
	$Z_3$ and $ Z_2 \times Z_2$ models respectively.}
\label{f:khlam}
\end{figure}


\begin{figure}[htbp!]
  \centering
  \includegraphics[width=0.45\textwidth]{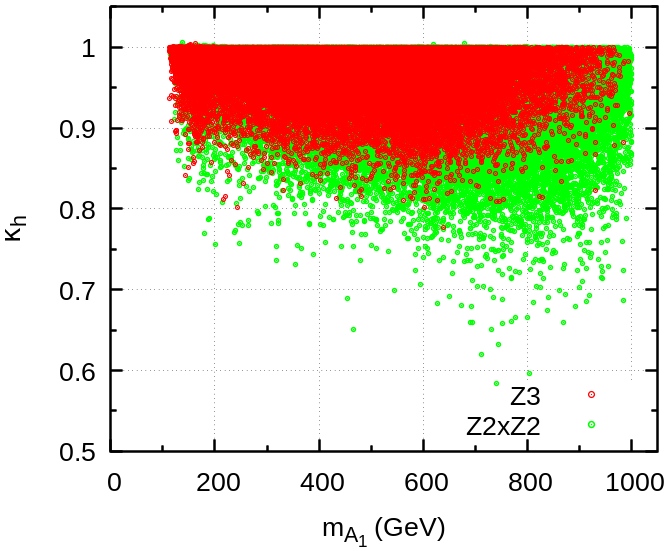}
  \includegraphics[width=0.45\textwidth]{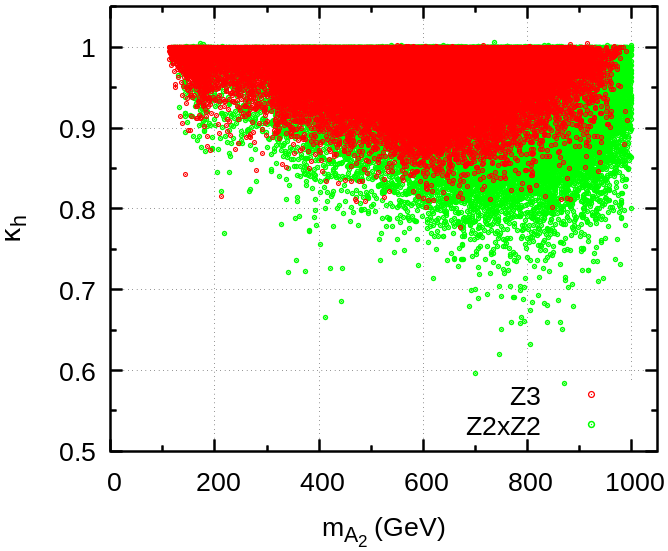}
  \caption{Points that pass through all the constraints are plotted in the
    $\kappa_h$ vs $m_{A1}, m_{A2}$ plane. The scanning is
    done assuming the Al-10\%  condition of \Eqn{e:Al-10}.
    The red and the green points correspond to the
    $Z_3$ and $ Z_2 \times Z_2$ models respectively.
     }
\label{f:khmass}
\end{figure}

Next we turn our attention to the trilinear Higgs self-coupling of the
following form:
\begin{eqnarray}
	{\mathscr L}_{hhh} = g_{hhh} h^3 \,.
\end{eqnarray}	
In the SM we have $g_{hhh}^{\rm SM} = -m_h^2/(2v)$.
Thus we define the following coupling modifier
\begin{eqnarray}
	\kappa_h = \frac{g_{hhh}}{g_{hhh}^{\rm SM}}
\end{eqnarray}
which is already being measured experimentally and some preliminary values
have been reported in Refs.~\cite{CMS:2022cpr,ATLAS:2021ifb}. We have checked that for both the
Type-Z models, $\kappa_h = 1$ in the alignment limit defined by
\Eqn{e:alignment}, as expected. Therefore we have to hope that the LHC Higgs data will
eventually settle for some nonstandard values away from exact alignment 
so that some distinguishing features can be found. To this end we recall
that the quartic parameters of \Eqn{e:VZ2Z3} mark the essential difference
between the two models. It should also be noted that in the limit
$\lambda_{10}^{(\prime)}, \lambda_{11}^{(\prime)}, \lambda_{12}^{(\prime)} = 0$, the
quartic part of the potential possesses a $U(1)\times U(1)$ symmetry
(independent from the $U(1)_Y$ hypercharge symmetry). Consequently,
$\lambda_{10}^{(\prime)}$, $\lambda_{11}^{(\prime)}$ and $\lambda_{12}^{(\prime)}$ are
the only quartic parameters that get involved in the expressions of the
pseudoscalar masses, $m_{A1}$ and $m_{A2}$. Keeping these in mind we
exhibit in Fig.~\ref{f:khlam} the scatter plot of the points that pass through
all the constraints in the $\kappa_h$ vs $\lambda_{10}^{(\prime)}, \lambda_{11}^{(\prime)}, \lambda_{12}^{(\prime)}$ plane. There we observe
that values of $\kappa_h$ in the ballpark $0.8$ or lower will definitely
favor the $Z_2\times Z_2$ scenario over the $Z_3$ version of Type-Z 3HDM.
To give these results a better physical context, in Fig.~\ref{f:khmass},
 we plot the same points in
the $\kappa_h$ vs pseudoscalar mass planes. This figure clearly indicates
that unlike the $Z_3$ model, the $Z_2\times Z_2$ model can still allow
$\kappa_h$ values as low as $0.7$.
In passing, we also note that values of $\kappa_h$ around $1.1$ or higher
 will disfavor both versions of Type-Z 3HDMs.

%
\begin{table}[htbp!]
  \centering
  \begin{tabular}{|c|c|c|c|c|c|c|}
    \hline
    & $m_{H1}$ & $m_{H2}$ & $m_{A1}$ &  $m_{A2}$ & $m_{C1}$ & $m_{C2}$ \\
    \hline\hline
    Benchmark 1 & 365  &    450  &  340 &    470 &  335 &   465  \\*[1mm]
    Benchmark 2 &  530 &  645 &  515 &  610 &   540 &     610   \\*[1mm]
    Benchmark 3 &  641  &  775 &  615 &     745 &    645  &   770 \\
    \hline
  \end{tabular}
  \caption{Benchmark values for the nonstandard masses (in GeV) used in Fig.~\ref{f:bench}.
     }
  \label{t:bench}
\end{table}

\begin{figure}[htbp!]
  \centering
  \includegraphics[width=0.44\textwidth]{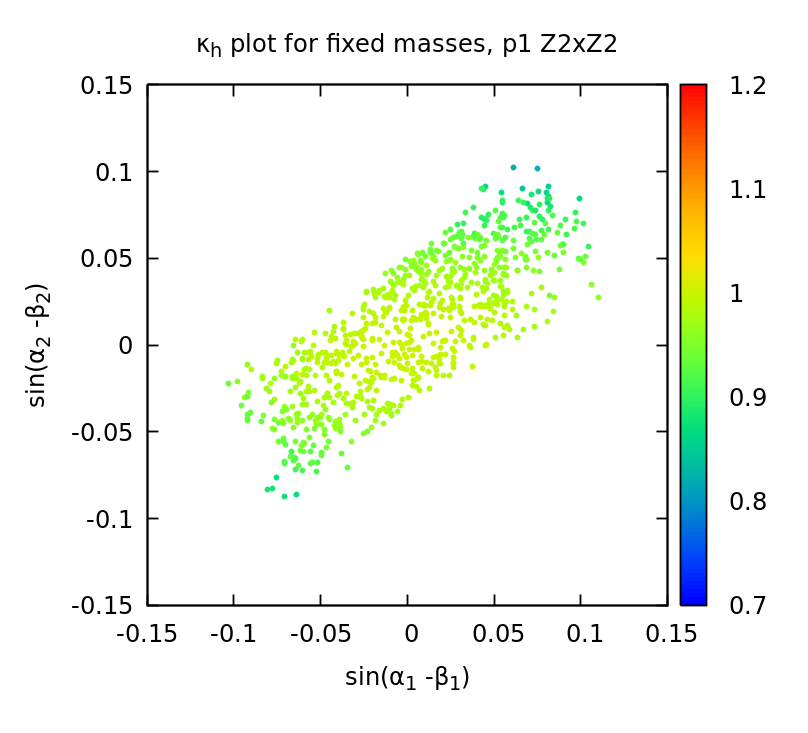} ~~
  \includegraphics[width=0.44\textwidth]{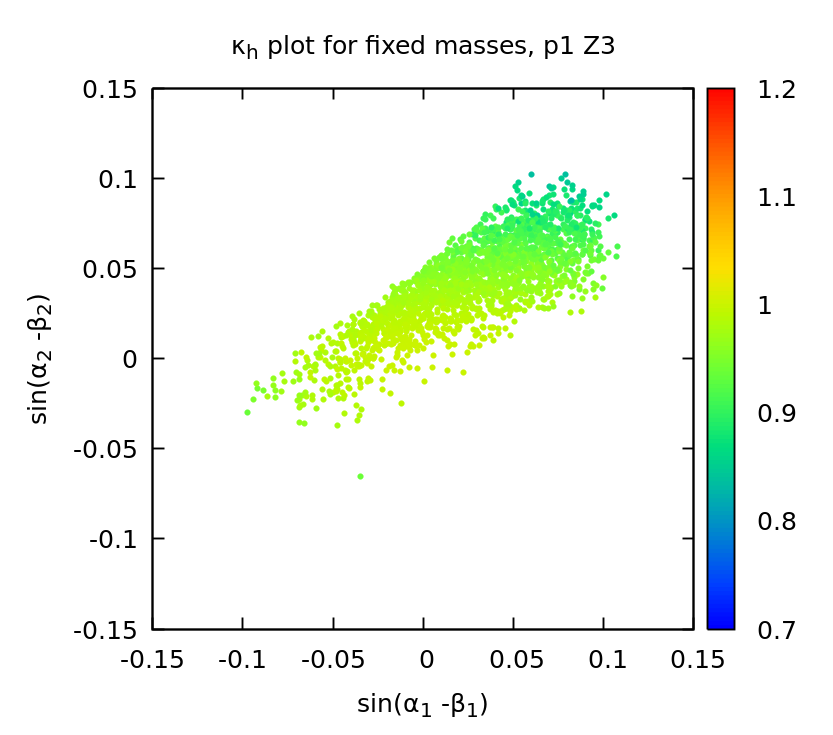} \\
  \includegraphics[width=0.44\textwidth]{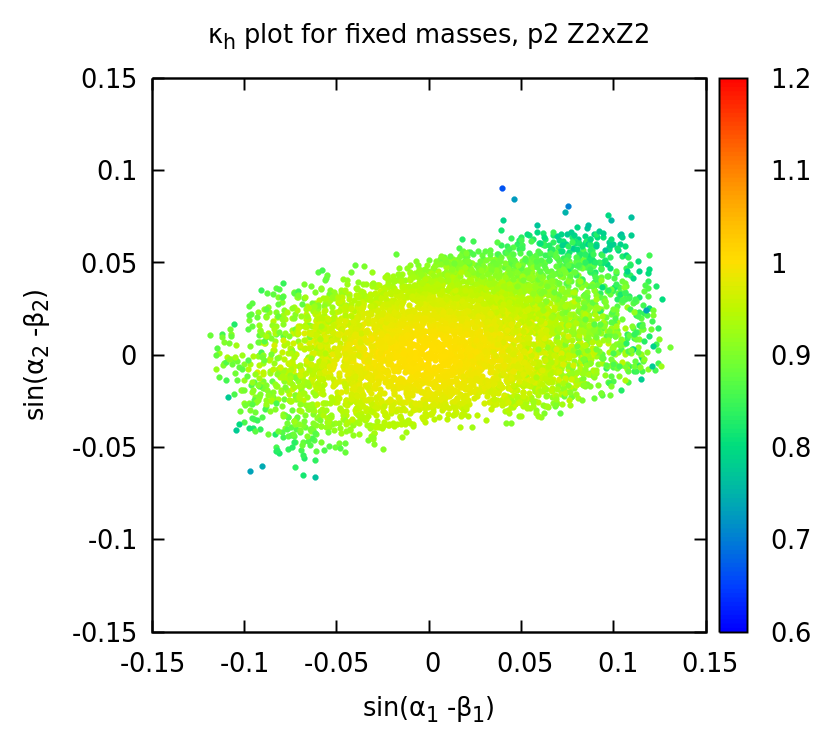} ~~
  \includegraphics[width=0.44\textwidth]{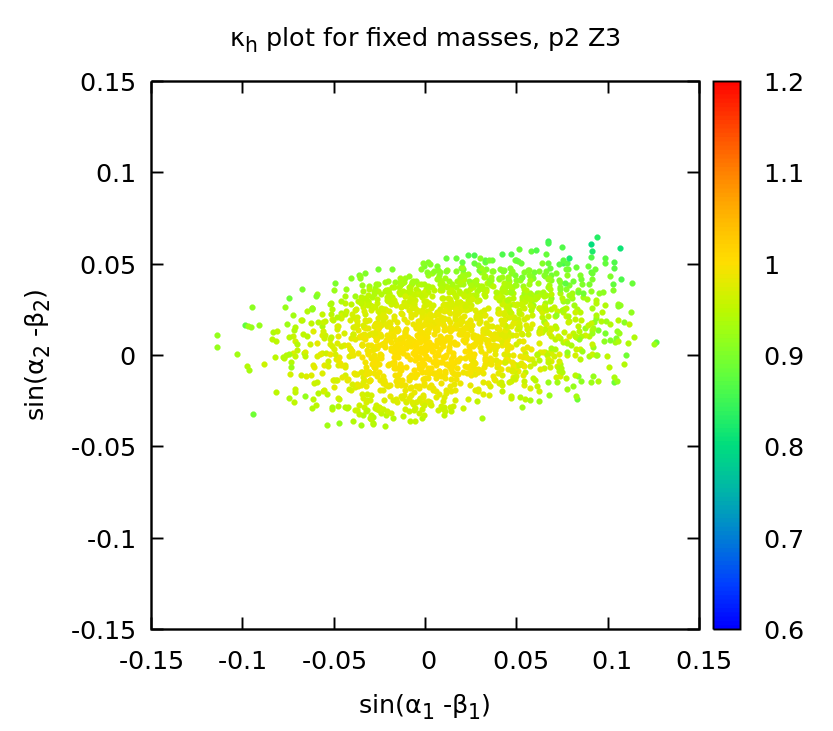} \\
  \includegraphics[width=0.44\textwidth]{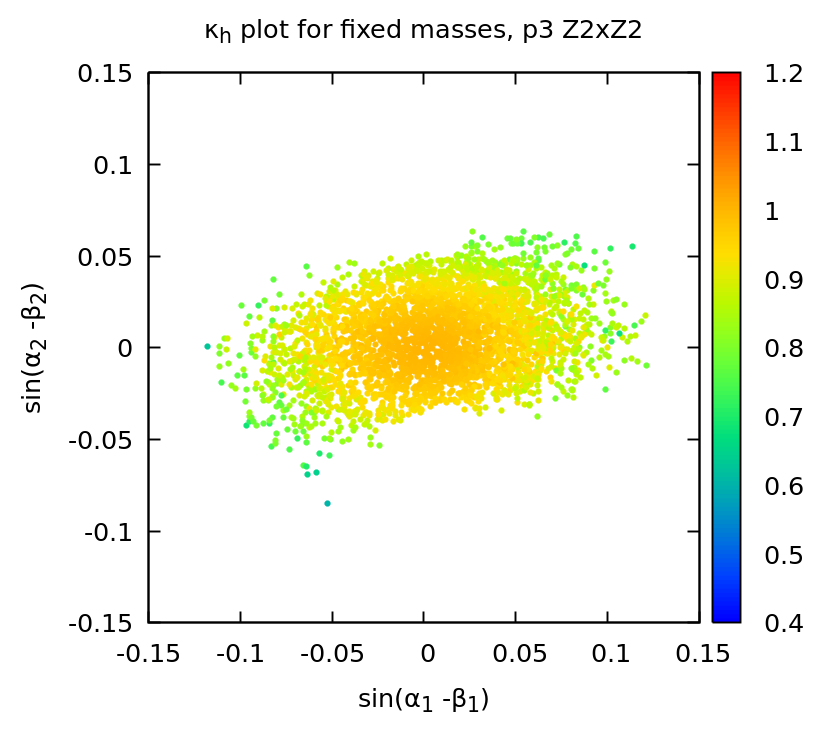} ~~
  \includegraphics[width=0.44\textwidth]{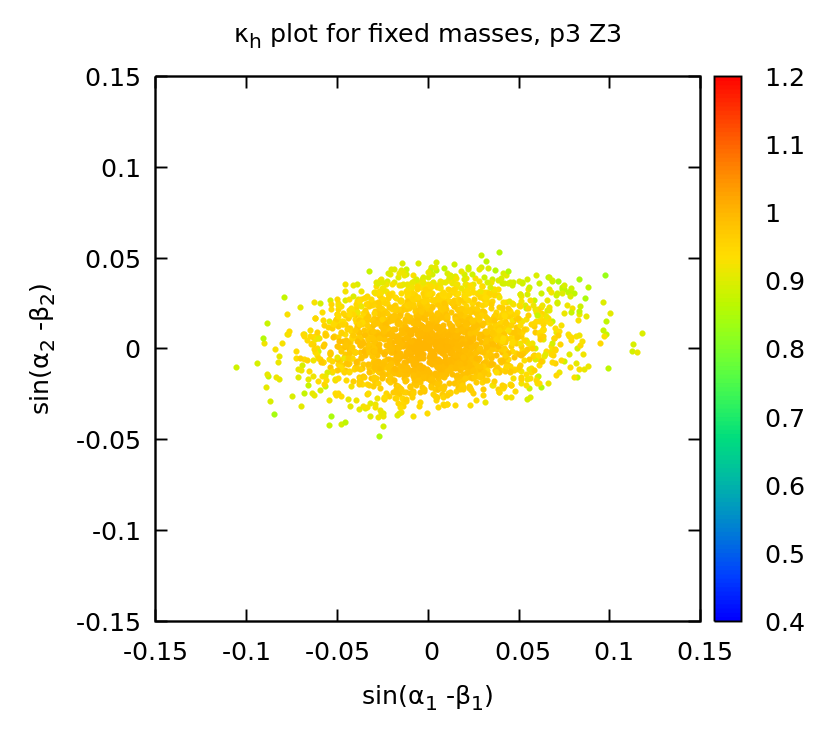}
  \caption{Plot in the $\sin{(\alpha_1-\beta_1)}$ vs
  $\sin{(\alpha_1-\beta_1)}$ plane for the benchmark values
(labeled appropriately) of Table~\ref{t:bench}. The
color bar associated with each plot marks the gradient of
values taken by $\kappa_h$. The plots in the left panel
correspond to the $Z_2\times Z_2$ model whereas the plots in
the right panel correspond to the $Z_3$ model.
Clearly, the distinguishability between the two models depends
on the benchmark point (mass region) chosen.
}
\label{f:bench}
\end{figure}

In a final and more optimistic effort, we presume that some nonstandard scalars have
already been observed and we try to ascertain whether, in view of the set of
nonstandard parameters, one of the Type-Z 3HDMs can be preferred over the other.
Our benchmark values for the nonstandard masses appear in Table~\ref{t:bench}.
The remaining parameters are scanned following \Eqn{e:Al-20}. 
For these benchmark values we have plotted all the points that pass through the
constraints in the $\sin(\alpha_1-\beta_1)$ vs $\sin(\alpha_2-\beta_2)$ plane.
The results have been displayed in Fig.~\ref{f:bench} where we have also color
coded the value of $\kappa_h$ for each point. There we can see that the points
span a relatively larger region for the $Z_2\times Z_2$ model. Therefore, if both
$\sin(\alpha_1-\beta_1)$ and $\sin(\alpha_2-\beta_2)$ are measured to be close to
$0.1$ along with $\kappa_h$ to be around $0.7$, then
it would definitely point towards the $Z_2\times Z_2$ model.
Thus, again, we have found that although we can find corners in the parameter space
that can isolate the $Z_2\times Z_2$ model, it seems to be very difficult to point
out exclusive features characterizing the $Z_3$ version of the Type-Z 3HDM.


\section{Summary}
\label{s:conclusions}
To summarize, we have studied the two common incarnations of Type-Z 3HDMs. One
of them employs a $Z_2\times Z_2$ symmetry while the other relies on a $Z_3$
symmetry. We point out that the difference between these two models is captured
by certain quartic terms in the scalar potential appearing in \Eqn{e:VZ2Z3}.
Then we proceed to uncover the effects of these quartic terms in creating
distinctions between the two Type-Z models.

In doing so we have performed exhaustive scans over the set of free parameters
in these models. Wherever possible, we have conveniently traded the Lagrangian
parameters in favor of the physical masses and mixings. Even then, when all the
relevant theoretical and experimental constraints are imposed, a completely
random scan generates very few output points that successfully negotiate all the
constraints. Therefore, we adopt a more strategic scanning procedure which involve
generating random points around a premeditated proximity of the `maximally
symmetric limit' defined by \Eqn{eq:max-sym}. In this way we have successfully
generated sufficient number of points to populate our plots.

For the plots, we were mainly interested in observables that involve the Higgs
self couplings. We have found that although $\mu_{\gamma\gamma}$ and
$\mu_{Z\gamma}$ are not the best discriminators, the trilinear Higgs 
self coupling modifier $(\kappa_h)$ has the potential to distinguish between
the two models. We have concluded that relatively lower values of $\kappa_h$
will favor the $Z_2\times Z_2$ version of Type-Z 3HDM. We have also emphasized
that some nonstandard physics need to be discovered in the LHC Higgs data
for us to be able to discriminate between the two Type-Z 3HDMs. Our study
underscores the importance of the ongoing effort to measure the trilinear
Higgs self coupling with increased precision.

\section*{Acknowledgments}
\noindent
	This work is supported in part by the Portuguese Funda\c{c}\~{a}o
para a Ci\^{e}ncia e Tecnologia\/ (FCT) under Contracts
CERN/FIS-PAR/0002/2021, CERN/FIS-PAR/0008/2019, UIDB/00777/2020, and UIDP/00777/2020\,;
these projects are partially funded through POCTI (FEDER),
	COMPETE, QREN, and the EU. The work of R. Boto is also supported
	by FCT with the PhD grant PRT/BD/152268/2021.
DD thanks the Science and Engineering Research Board,
India for financial support  through grant number CRG/2022/000565.


\appendix

\section{\label{app:impact}Impact of a wider search}

In order to assess the need for a search of points close to the alignment limit of
\Eqs{e:alignment}{eq:max-sym},
we redo Table \ref{tab:k10}, now with the looser bounds
\begin{equation}
\label{e:Al-50}
\frac{\alpha_1}{\beta_1},\ 
\frac{\alpha_2}{\beta_2},\ 
\frac{\gamma_2}{\gamma_1},\ 
\frac{-\alpha_3}{\gamma_1},\ 
\frac{m_{A1}}{m_{H1}},\ 
\frac{m_{C1}}{m_{H1}},\ 
\frac{m_{A2}}{m_{H2}},\ 
\frac{m_{C2}}{m_{H2}}\, \in\, [0.5,1.5]\, .
\ \ \ \textbf{(Al-50\%)}
\end{equation}
\begin{table}[htbp!]
\centering
\makebox[0pt][c]{\parbox{0.97\textwidth}{%
    \begin{minipage}[b]{0.47\textwidth}\centering
      \begin{tabular}{|c|c|c|c|c|}\hline
        \multicolumn{5}{|c|}{$ Z_2 \times  Z_2$\ (Al-50\%)  }\\\hline\hline
        Check    &     N       &  Y     & 100*p   & 100*$\delta_p$\\\hline\hline
        STU   &   500000  &     46179   &      9.236  &     0.045\\*[1mm]
        BFB  &   5000000  &    277802  &     5.556 &      0.011\\*[1mm]
        Unitarity   &   500000  &      3397  &     0.679 &      0.012\\*[1mm]
        $b \rightarrow s \gamma$   &    50000  &     19340  &    38.680  &     0.328\\*[1mm]
        $\mu$'s   &    50000   &    217   & 0.434    &    0.077 \\\hline
      \end{tabular}
    \end{minipage}
    \hfill
    \begin{minipage}[b]{0.47\textwidth}\centering
      \begin{tabular}{|c|c|c|c|c|}\hline
        \multicolumn{5}{|c|}{$ Z_3$\  (Al-50\%) }\\\hline\hline
        Check    &     N       &  Y     & 100*p   & 100*$\delta_p$\\\hline\hline
        STU   &   500000  &     46296  &     9.259  &     0.045\\*[1mm]
        BFB  &   5000000  &     25673  &     0.513  &     0.003\\*[1mm]
        Unitarity  &    500000  &      2639  &     0.528  &     0.010\\*[1mm]
        $b \rightarrow s \gamma$  &     50000  &     19193  &    38.386  &     0.326\\*[1mm]
        $\mu$'s  &     50000  &    178   & 0.356     & 0.069    \\\hline
      \end{tabular}
    \end{minipage}
  }}
\caption{Impact of individual constraints for the two Type-Z models
  while the scanning is done following \Eqn{e:Al-50}.}
\label{tab:k16}
\end{table}
Comparing Table~\ref{tab:k10} with Table~\ref{tab:k16},
we notice that, away from alignment, the unitarity and $\mu$ constraints cut
most of the allowed parameter space.

\bibliographystyle{JHEP}
\bibliography{BFB}

\end{document}